\documentclass[lettersize,conference]{IEEEtran}
\usepackage{amsmath,amsfonts}
\usepackage[ruled,linesnumbered]{algorithm2e}
\usepackage[caption=false,font=footnotesize,labelfont=rm,textfont=rm]{subfig}
\usepackage{makecell}
\usepackage{stfloats}
\usepackage{ragged2e}
\usepackage{url}
\usepackage{verbatim}
\usepackage{graphicx}
\usepackage{balance}
\usepackage{cite}
\usepackage{multirow}
\usepackage{enumitem}
\usepackage{tikz}
\usetikzlibrary{shapes, arrows.meta, positioning}
\usepackage{booktabs}
\usepackage{lipsum}
\usepackage{amssymb}
\usepackage[colorlinks,linkcolor=blue,anchorcolor=blue,citecolor=blue]{hyperref}

\newcommand{\myrefeq}[1]{(\ref{#1})}
\newcommand{\myreffig}[1]{Fig. \ref{#1}}
\newcommand{\myreftable}[1]{Table \ref{#1}}
\newcommand{\myrefalgo}[1]{Algorithm \ref{#1}}

\begin{document}

	\title{Distributed MoE-based Uplink Detection for Cell-Free Communication Systems}

	\author{
		Le Zhao$^*$, Xuesong Pan$^*$, Xinyi Wang$^{*\dagger}$, Zhong Zheng$^{*\dagger }$, Zesong Fei$^{*\dagger}$\\
		$^*$ School of Information and Electronics, Beijing Institute of Technology, Beijing 100081, China\\
		$^\dagger$ National Key Laboratory of Science and Technology on Space-Born Intelligent Information Processing, \\Beijing Institute of Technology, Beijing 100081, China\\
		Contact Email: tobin\_bit@icloud.com
	}

	\maketitle

	\begin{abstract}
		Cell-free Massive multiple input and multiple output (MIMO) is recognized as a key technology for beyond-5G networks, where distributed access points (APs) jointly serve user equipments (UEs) to address the inherent inter-cell interference issue inherent in cellular systems. While conventional distributed signal detection methods offer a practical balance between performance and fronthaul load, they are fundamentally limited by linear processing constraints. In this paper, we propose a novel deep learning based uplink detection framework by introducing the distributed mixture of experts detection network (DMoE-DetNet). In this architecture, each AP acts as a local expert employing convolutional neural networks (CNNs) for non-linear feature extraction, and transmits the local minimum mean square error (MMSE) detection results and statistical channel information to the central processing unit (CPU). In the CPU, an attention-based encoder module captures complex spatio-temporal dependencies among users for global feature fusion, with a gating network at the central processor dynamically weighting the contributions from different APs. At last, a linear detector outputs the symbol probability. Simulation results demonstrate that the proposed DMoE-DetNet significantly outperforms conventional linear processing based cell-free signal detection methods in terms of symbol error rate, showcasing the potential of artificial intelligence-enabled communication systems.
	\end{abstract}


	\section{Introduction}

	Global mobile user equipment (UE) connections are projected to reach 7.49 billion by 2025. Multiple-input multiple-output (MIMO) technology has been fundamental to 5G and beyond-5G networks \cite{Marzetta2010a}, delivering over 10× spectral efficiency gains compared to legacy systems through base station hardware upgrades rather than new site deployments \cite{bjornson2017massive, DetectorYao2022}.

	In the uplink MIMO systems, efficiently and accurately receiving and processing signals transmitted by UEs has always been a core challenge in communication systems. While linear signal detection techniques such as s minimum mean-squared error (MMSE) and zero forcing (ZF) combining, have been widely adopted for their reasonable performance and acceptable implementation complexity \cite{Altamirano2019}, their performance are fundamentally limited by linear processing constraints. Artificial intelligence (AI) is playing a promising role in MIMO signal detection to overcome the limitations of these conventional MIMO detectors \cite{Nguyen2023AI-MIMO}. Several AI-based approaches, including DetNet \cite{Samuel2019}, OAMP-Net \cite{He2020}, and LISA \cite{Sun2020} have demonstrated promising results by leveraging deep learning for MIMO detection. Nevertheless, the mentioned methods are limited by the inter-cell interference and network coverage since a non-cooperative cellular network where each base station serves an exclusive set of UEs.

	Inspired by cell-free architecture proposed for the scenario where a large number of access points (APs) and central processing unit (CPU) are jointly serving UEs with no cell boundaries, the distributed signal detection has emerged as a promising cellular paradigm. Early studies such as \cite{Ngo2017b} focused on configurations where distributed APs are connected to a CPU via fronthaul links, forwarding both channel state information (CSI) and received signals to facilitate centralized detection. This architecture, denoted as a fully centralized processing framework, however, introduces substantial fronthaul overhead and increases processing latency.
	To overcome the main drawback of the centralized detection method, several distributed schemes such as local CSI utilization for intermediate computations \cite{Rompaey2021CFD}, expectation propagation with inter-module iteration\cite{He2021globalcomCDF}, and partial MMSE reception where users are served by partial nodes \cite{Nayebi2016CDF} were proposed.

	Among the multiple low-overhead distributed detection schemes, large-scale fading decoding (LSFD) \cite{Adhikary2017tcomLSFD, Xuesong2024TWC} has attracted considerable attention. LSFD operates in two distinct stages: local detection at the APs and weighted combining at the CPU, where the combining weights are determined based on channel large-scale channel fading coefficients. In \cite{Nayebi2016CDF}, the authors derived closed-form expressions for the optimal combining weights under the assumption that matched filters (MF) are used at the local receivers. while \cite{Zhang2021tcomZFLSFD} and \cite{Bjornson2019a} extended to ZF and MMSE detectors incorporating channel estimation errors. However, the optimization of LSFD combining weights relies on the expectation-maximization process and statistical channel assumptions, leading to performance degradation under complex deterministic channel conditions. Moreover, practical deployment requires the CPU to periodically acquiring CSI from UEs to APs to observe channel statistics and update the weights, which introduces additional overhead.

	To overcome these limitations and realize the full potential of distributed processing, we propose the distributed mixture of experts detection network (DMoE-DetNet) framework. In this structure, each AP acts as a local expert, leveraging instantaneous CSI (ICSI) for precise signal processing, while a gating network at the CPU uses statistical CSI (SCSI) for adaptive expert fusion. Furthermore, a Transformer encoder is utilized to capture complex spatio-temporal dependencies that are challenging for conventional linear processors. Once well-trained, the proposed framework operates using only the detected signals from APs and SCSI, which can be acquired in prior via, e.g., channel knowledge map (CKM) \cite{Levie_2021_TWC, Le3DRadiodiff2025wcl, ZengYong2024TWCCKMBeamforming}.

	The rest of this paper is organized as follows. Section~\ref{sec:system_model} presents the system model and reviews conventional distributed reception schemes. Section~\ref{sec:proposed_method} details the proposed DMoE-DetNet architecture. Simulation results and discussions are provided in Section~\ref{sec:simulation}, and Section~\ref{sec:conclusion} concludes the paper.

	\section{System Model and Conventional Detection}
	\label{sec:system_model}

	\subsection{Cell-Free MIMO System Model}

	\begin{figure}[!t]
		\centering
		\includegraphics[width=0.9\columnwidth]{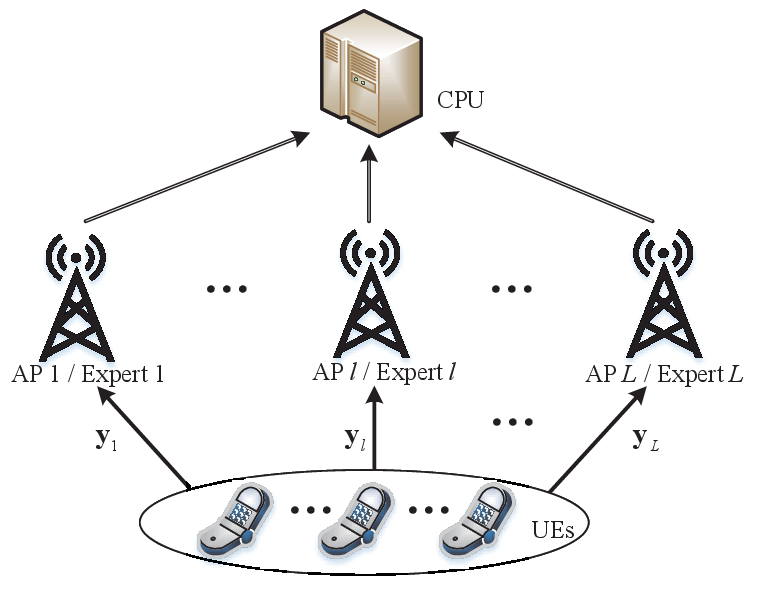} 
		\caption{Illustration of the cell-free MIMO uplink system.}
		\label{fig:system_model}
	\end{figure}

	As illustrated in \myreffig{fig:system_model}, we consider a cell-free MIMO uplink system where $K$ single-antenna UEs are served by $L$ distributed APs, each equipped with $N$ receive antennas. All APs are connected to a CPU via fronthaul links. The channel between UE $k$ and AP $l$ is modeled as spatially correlated Rayleigh fading:
	\begin{equation}
		\mathbf{h}_{kl} \sim \mathcal{CN}(\mathbf{0}, \mathbf{R}_{kl}),
	\end{equation}
	where $\mathbf{R}_{kl} \in \mathbb{C}^{N \times N}$ is the spatial correlation matrix, and $\beta_{kl} \triangleq \mathrm{tr}(\mathbf{R}_{kl})/N$ represents the large-scale fading coefficient.

	During uplink data transmission, each AP first estimates the local CSI from UEs and performs MMSE signal detection based on estimated CSI. The CPU subsequently combines the locally detected signals from all APs through weighted fusion to enhance the overall detection accuracy.

	The received signal at AP $l$ is modeled as:
	\begin{equation}
		\mathbf{y}_l = \sum_{k=1}^{K} p_k \mathbf{h}_{kl} \mathcal{M}(s_k) + \mathbf{n}_l,
	\end{equation}
	where $ \mathcal{M}(\cdot) $ denotes the modulation mapping function, $ s_k \in \mathcal{S} $ is the transmitted symbol from UE $ k $ with $ \mathbb{E}[|s_k|^2] = 1 $, and $ \mathbf{n}_l \sim \mathcal{CN}(\mathbf{0}, \sigma_n^2 \mathbf{I}_N) $ represents the additive white Gaussian noise at AP $ l $.

	To estimate the transmitted symbols, an MMSE detector is applied to minimize the mean-square error. Using the estimated channel $ \hat{\mathbf{h}}_{kl} $, the detection matrix for AP $ l $ is given by
	\begin{align}
		\mathbf{v}_{l}^{\text{MMSE}} = \left( \hat{\mathbf{h}}_{kl}^H \hat{\mathbf{h}}_{kl} + 	\sigma_n^2 \mathbf{I}_{N} \right)^{-1} \hat{\mathbf{h}}_{kl}^H,
	\end{align}
	and the corresponding symbol estimate is obtained as:
	\begin{align}
		\hat{\mathbf{s}}_{l}^{\text{MMSE}} = \mathbf{v}_l^{\text{MMSE}} \mathbf{y}_l.
	\end{align}
	However, in multi-cell scenarios, inter-cell interference severely degrades the accuracy of the channel estimate $ \hat{\mathbf{h}}_{kl} $, which in turn constrains the detection performance at the $l$-th AP.

	\subsection{Fully Centralized Processing}
	To overcome the performance limitations caused by inter-cell interference in local detection, the fully centralized processing framework is first considered \cite{Ngo2017b}. In this architecture, all APs forward their received signals to the CPU for joint processing, thereby mitigating the impact of local channel estimation errors.
	The aggregate received signal across all APs is given by:
	\begin{equation}
		\mathbf{y} = \mathbf{H}\mathbf{x} + \mathbf{n},
	\end{equation}
	where
	\begin{equation}
		\mathbf{y} = \begin{bmatrix} \mathbf{y}_1 \\ \vdots \\ \mathbf{y}_L \end{bmatrix},
		\mathbf{H} = \begin{bmatrix} \mathbf{H}_1 \\ \vdots \\ \mathbf{H}_L \end{bmatrix},
		\mathbf{x} = \begin{bmatrix} p_1\mathcal{M}(s_1) \\ \vdots \\ p_K\mathcal{M}(s_K) \end{bmatrix},
		\mathbf{n} = \begin{bmatrix} \mathbf{n}_1 \\ \vdots \\ \mathbf{n}_L \end{bmatrix}.
	\end{equation}
	Here, \(\mathbf{H}_l = [\mathbf{h}_{1l}, \dots, \mathbf{h}_{Kl}] \in \mathbb{C}^{N \times K}\) denotes the channel matrix at AP \(l\), and \(\mathbf{n} \sim \mathcal{CN}(\mathbf{0}, \sigma_n^2 \mathbf{I}_{LN})\) is the composite noise vector.

	Equivalently, the signal model can be expressed as:
	\begin{equation}
		\mathbf{y} = \sum_{i=1}^{K} \mathbf{h}_i s_i + \mathbf{n},
	\end{equation}
	where \(\mathbf{h}_i = [\mathbf{h}_{i1}^T, \ldots, \mathbf{h}_{iL}^T]^T \in \mathbb{C}^{LN}\) represents the aggregated channel vector from UE \(i\) to all APs.

	The CPU performs MMSE detection to estimate the transmitted symbols:
	\begin{equation}
		\hat{s}_k = \mathbf{v}_k^H \mathbf{y},
	\end{equation}
	with the optimal combining vector formulated as:
	\begin{equation}\label{eq9}
		\mathbf{v}_k = p_k \left( \sum_{i=1}^{K} p_i \left( {\mathbf{h}}_i {\mathbf{h}}_i^H \right) + \sigma_n^2 \mathbf{I}_{LN} \right)^{-1} \hat{\mathbf{h}}_k.
	\end{equation}
	In this paper, we assume perfect CSI is available.

	While this centralized approach achieves theoretically optimal performance, it requires all APs to transmit their complete received signals to the CPU, resulting in prohibitive fronthaul overhead that limits practical implementation in large-scale systems.

	\subsection{Distributed Cell-Free MIMO Detection}

	\subsubsection{Local Processing with LSFD Combining \cite{Bjornson2019a}}

	In this distributed approach, each AP \(l\) first computes local estimates of the transmitted symbols using linear combiners. For UE \(k\), the local estimate at AP \(l\) is obtained as
	\begin{equation}
		\check{s}_{kl} = \mathbf{v}_{kl}^H \mathbf{y}_l,
	\end{equation}
	where \(\mathbf{v}_{kl} \in \mathbb{C}^N\) is the local combining vector.

	The local MMSE combiner at AP \(l\) for UE \(k\) is given by:
	\begin{equation}\label{MMSE}
		\mathbf{v}_{kl} = p_k \left( \sum_{i=1}^{K} p_i \left( \hat{\mathbf{h}}_{il} \hat{\mathbf{h}}_{il}^H + \mathbf{C}_{il} \right) + \sigma_n^2 \mathbf{I}_N \right)^{-1} \hat{\mathbf{h}}_{kl}.
	\end{equation}
	These local estimates are then forwarded to the CPU, which performs LSFD to produce the final symbol estimate.

	In particular, the optimal LSFD weights are derived based on channel statistics as
	\begin{equation}
		\mathbf{a}_k = \left( \sum_{i=1}^{K} p_i \mathbb{E}\{\mathbf{g}_{ki} \mathbf{g}_{ki}^H\} + \sigma_n^2 \mathbf{D}_k \right)^{-1} \mathbb{E}\{\mathbf{g}_{kk}\},
	\end{equation}
	where $
	\mathbf{g}_{ki} = \left[ \mathbf{v}_{k1}^H \mathbf{h}_{i1}, \ldots, \mathbf{v}_{kL}^H \mathbf{h}_{iL} \right]^T$, and $
	\mathbf{D}_k = \mathrm{diag}\left( \mathbb{E}\{\|\mathbf{v}_{k1}\|^2\}, \ldots, \mathbb{E}\{\|\mathbf{v}_{kL}\|^2\} \right)$.
	The final LSFD-based estimated signal is calculated as
	\begin{equation}
		\hat{s}_k = \sum_{l=1}^{L} a_{kl}^* \check{s}_{kl}.
	\end{equation}

	This method achieves a favorable balance between detection performance and fronthaul overhead by exploiting statistical channel information.

	\subsubsection{Local Processing with Averaging Combining}

	A simplified fusing scheme for CPU is to execute equal weighting for all APs as
	\begin{equation}
		\hat{s}_k = \frac{1}{L} \sum_{l=1}^{L} \check{s}_{kl}.
	\end{equation}

	While this approach minimizes fronthaul signaling and computational complexity at the CPU compared with LSFD combining, it yields suboptimal performance due to the non-adaptive averaging of local estimates, particularly in heterogeneous channel conditions.

	\begin{figure}[!t]
		\centering
		\includegraphics[width=1\columnwidth]{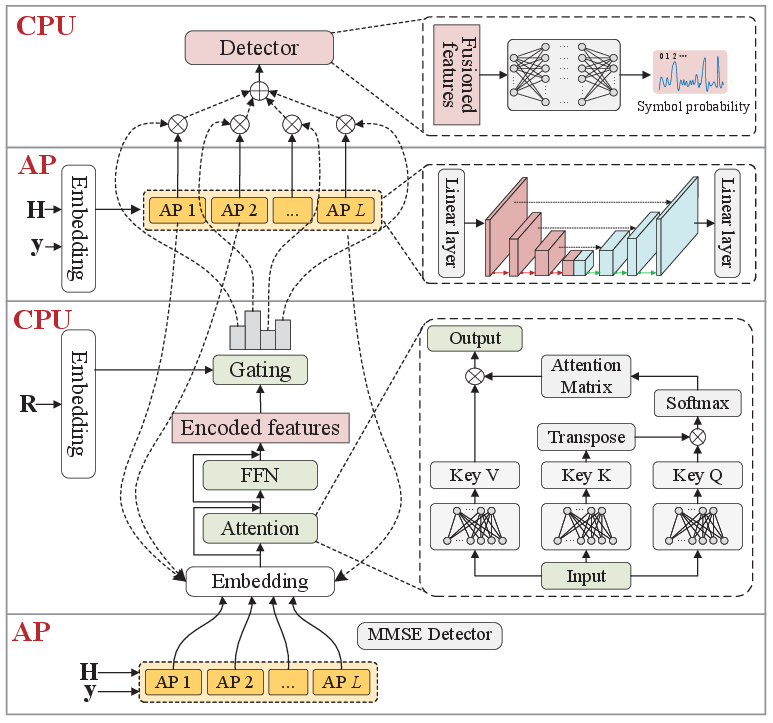} 
		\caption{Illustration of the cell-free MIMO uplink system.}
		\label{fig:structure}
	\end{figure}

	\section{Proposed DMoE-DetNet}
	\label{sec:proposed_method}

	In this section, we present the proposed DMoE-DetNet framework. While ICSI enables precise local signal processing, SCSI provides the statistical context for intelligent fusion. Our architecture explicitly leverages this distinction through a novel division of labor.

	\subsection{Core Design Principle}

	The DMoE-DetNet embodies a principled approach to distributed detection by assigning distinct roles to different types of channel information:

	\begin{itemize}
		\item {ICSI for Expert Networks:} Each AP's expert network utilizes ICSI estimation $\{\hat{\mathbf{h}}_{kl}\}$ to perform sophisticated, non-linear local processing that adapts to fast-fading variations.

		\item {SCSI for Gating Network:} The central gating network employs SCSI $\{\mathbf{R}_{kl}\}$ to compute fusion weights that reflect long-term channel statistics and network geometry.
	\end{itemize}

	This design mirrors and generalizes conventional distributed reception: the expert networks learn non-linear counterparts to local combining, while the gating network learns optimized replacements for LSFD weights. However, unlike conventional approaches constrained by linear processing, DMoE-DetNet leverages deep learning to discover more effective non-linear strategies.

	\subsection{Architecture Components}
	As shown in \myreffig{fig:structure}, the components of DMoE-DetNet specifically consist of the following components:

	\subsubsection{MMSE detection}

	Upon receiving the signals, each AP firstly performs MMSE detection to obtain a preliminary estimate $\hat{\mathbf{s}}_{\rm MMSE} = [\hat{\mathbf{s}}_1, \dots, \hat{\mathbf{s}}_L]$, as predicted in \myrefeq{MMSE}. These local estimates are then transmitted to the CPU, where a neural network captures complex spatio-temporal dependencies for further refinement.

	\subsubsection{Distributed Experts}

	Each AP $l$ hosts an expert network that processes locally available information
	\begin{equation}
		\mathbf{F}_l = \rm{Expert}_l\left(\rm{Concat}\left(\mathbf{y}_l, \Re(\hat{\mathbf{H}}), \Im(\hat{\mathbf{H}})\right)\right),
	\end{equation}
	where \(\hat{\mathbf{H}}_l\) denotes the instantaneous CSI (ICSI) of all UEs observed at AP \(l\). The output \(\mathbf{F}_l \in \mathbb{C}^{K \times D_{\mathbf{F}}}\) is a fused representation, with \(D_{\mathbf{F}}\) being the dimension of the expert’s final linear layer. Here, \(\Re(\cdot)\) and \(\Im(\cdot)\) extract the real and imaginary parts, respectively.

	\subsubsection{Feature Encoder}
	With input from MMSE detection at APs, CPU employs a Transformer encoder to captures complex spatio-temporal dependencies through self-attention, ultimately producing the detected symbols. The global encoded feature can be expressed as
	\begin{align}
		\tilde{\mathbf{s}}_{\rm global} = {\rm Transformer}(\hat{\mathbf{s}}_{\rm MMSE}, \mathbf{F}),
	\end{align}
	where $\mathbf{F} = [\mathbf{F}_1, \dots, \mathbf{F}_L]\in \mathbb{C}^{L \times K \times D_{\mathbf{F}}}$.

	\subsubsection{Intelligent Gating Network}

	The gating network at the CPU computes attention weights based on SCSI
	\begin{align}
		\mathbf{a}_l = \frac{\exp\left(\rm{Gate}\left( \{\mathbf{R}_{kl}\}, \tilde{\mathbf{s}}_{\rm global} \right)\right)}{\sum_{i=1}^{K}\exp\left(\rm{Gate}\left( \{\mathbf{R}_{kl}\}, \tilde{\mathbf{s}}_{\rm global} \right)\right)},
	\end{align}
	where $\mathbf{a}_l \in \mathbb{C}^{K \times 1}$ is the optimized LSFD weight of $l$-th AP for detect signal from $K$ UEs, the weight matrix is denoted as $\mathbf{A} = [\mathbf{a}_1, \dots, \mathbf{a}_L]^T \in \mathbb{C}^{L \times K}$.
	Note that fusing $\mathbf{R}_{kl}$ and $\tilde{\mathbf{s}}_{\rm global}$ enables dynamic, statistics-aware fusion that prioritizes contributions from APs with favorable long-term channel conditions.

	\subsubsection{Signal Detector}
	The fused vector—obtained by weighting and summing the outputs of the experts according to weight matrix $\mathbf{A}$—is passed through a linear fully-connected layer to output the probability of each symbol in the symbol set $\mathcal{S}$. This process can be expressed as:
	\begin{align}\label{loss}
		\mathcal{P}(\hat{\mathbf{s}}) = \mathrm{Linear}\left( \sum_{l=1}^{L} ( \mathbf{a}_l \cdot \mathbf{F}_l ) \right),
	\end{align}
	where $\mathbf{a}_l \in \mathbb{C}^{1 \times K}$ denotes the $l$-th row vector of $\mathbf{A}$. The detected signal vector $\hat{\mathbf{s}}$ is obtained from the output of this layer.

	During model training, the loss function measures the error between the predicted distribution in \myrefeq{loss} and the true symbol distribution $\mathcal{P}(\mathbf{s})$. The complete workflow of the proposed DMoE-DetNet detector is summarized in \myrefalgo{algo1}.

	\vspace{0.1in}
	\begin{algorithm}[!t]
		\label{algo1}
		\caption{DMoE-DetNet based Signal Detection}
		\KwIn{Received signals $\{\mathbf{y}_l\}$, ICSI $\{\hat{\mathbf{H}}_{kl}\}$, SCSI $\{\mathbf{R}_{kl}\}$}
		\KwOut{Detected symbols $\{\hat{s}_k\}$}
		\For{each AP $l = 1$ to $L$}{
			Estimate the local signal: $
			\hat{\mathbf{s}}_{\rm {MMSE}} = (\mathbf{H}^H \mathbf{H} + \sigma_n^2 \mathbf{I}_{N})^{-1} \mathbf{H}^H \mathbf{y}$\;
			Distributed experts processing: $\mathbf{F}_l = \rm{Expert}_l\left(\rm{Concat}\left(\mathbf{y}_l, \Re(\hat{\mathbf{H}}), \Im(\hat{\mathbf{H}})\right)\right)$\;
		}
		Encode the received signal in CPU: $\tilde{\mathbf{s}}_{\rm global} = {\rm Transformer}(\hat{\mathbf{s}}_{\rm MMSE})$\;
		Compute gating weights $\mathbf{A}$ based on \myrefeq{gating}\;
		Estimate the symbol probability distribution $\mathcal{P}(\hat{\mathbf{s}})$ based on \myrefeq{loss}\;
		\Return{Detected symbols $\{\hat{s}_k\}$}
	\end{algorithm}

	\section{Simulation Results Discussion}
	\label{sec:simulation}
	In this section, we evaluate the symbol error rate (SER) performance of the proposed DMoE-DetNet and compare it with several benchmarks. In particular, Rayleigh fading is considered for the channels between UEs and APs equipped with $N$ antennas. $\{4, 16\}-$QAM modulation is adopted for uplink transmission.

	\subsection{Data Generation}

	The channel realizations are generated using a spatially correlated Rayleigh fading model. The large-scale fading coefficients, incorporating both path loss and spatial correlation, are embedded in the covariance matrix $\mathbf{R}_{kl}$ for each UE $k$ and AP $l$ links. This matrix is constructed as $\mathbf{R}_{kl} = \text{PL}_{kl} \cdot \mathbf{\Phi}_{kl}$, where $\text{PL}_{kl}$ is the distance-based path loss, and $\mathbf{\Phi}_{kl} \in \mathbb{C}^{N \times N}$ is a normalized covariance matrix defining the spatial correlation. The instantaneous channel vector $\mathbf{h}_{kl}(t)$ is then generated as $\mathbf{h}_{kl}(t) = \mathbf{R}_{kl}^{1/2} \mathbf{w}(t)$, where $\mathbf{w}(t) \sim \mathcal{CN}(\mathbf{0}, \mathbf{I}_{N})$.
	This approach ensures the deterministic channel realizations to exhibit consistent large-scale statistics across different network snapshots, providing a deterministic foundation for evaluating detector performance.
	Note that perfect channel estimation is assumed at the APs, allowing the channel error terms in \myrefeq{eq9} and \myrefeq{MMSE} to be set to $\mathbf{0}$.

	For model training, we generate $10^5$ distinct large-scale fading matrices $\mathbf{R}_{kl}$ for each combination of $K$ UEs and $L$ APs. Each such matrix is associated with $N_{\rm seq}=14$ independent transmitted symbols $s_k$ within a single frame. The test dataset consists of $10^4$ independently generated large-scale matrices and their corresponding small-scale channel samples.


	\subsection{Model Configuration}
	The detailed configuration and parameter distribution of the proposed DMoE-DetNet are summarized in \myreftable{tab:model_config}.
	\vspace{-0.2cm}
	\begin{table}[!h]
		\renewcommand{\arraystretch}{0.7}
		\centering
		\setlength{\tabcolsep}{0.7mm}
		\caption{Model Architecture Summary}
		\label{tab:model_config}
		\begin{tabular}{l|l}
			\toprule[1.1pt]
			\textbf{Module} & \textbf{Core Components} \\
			\midrule
			\textbf{Input embedding} & \texttt{x\_embedding}: Linear($2 \times L$, 256) \\
			\midrule
			\textbf{Encoder} & \texttt{wq/wk/wv}: Linear(256, 256) \\
			& \texttt{dense}: Linear(256, 256) \\
			& 8 heads, dropout=0.1 \\
			\midrule
			\textbf{Gating network} & \texttt{R\_encoder}: Linear($2 \times N^2$, 128) + ReLU \\
			& \texttt{s\_proj}: Linear(256, 128) \\
			& \texttt{fusion}: Linear(256, 128) + ReLU + Linear(128, 1) \\
			\midrule
			\textbf{Expert} & \texttt{cnn2d}: Conv2d($256+2N$, 64, 3$\times$3) + ReLU \\
			& \quad Conv2d(64, 64, 3$\times$3) + ReLU \\
			& \quad Conv2d(64, 256, 1$\times$1) \\
			& $L$ experts in parallel \\
			\midrule
			\textbf{Detector} & \texttt{detectionHead}: Linear(256, $|\mathcal{S}|$) \\
			\bottomrule[1.1pt]
		\end{tabular}
	\end{table}

	The computaional complexity of DMoE-DetNet is primarily determined by the input dimensions sequence length $N_{seq}$, number of UEs $K$, number of APs $L$, and number of antennas $N$. Overall, the complexity is expressed as $O(L K N^2 + N_{seq} K^2 + N_{seq} K L N)$.

	\subsection{Comparison and Discussion}

	\begin{figure}[!t]
		\centering
		\includegraphics[width=0.955\columnwidth]{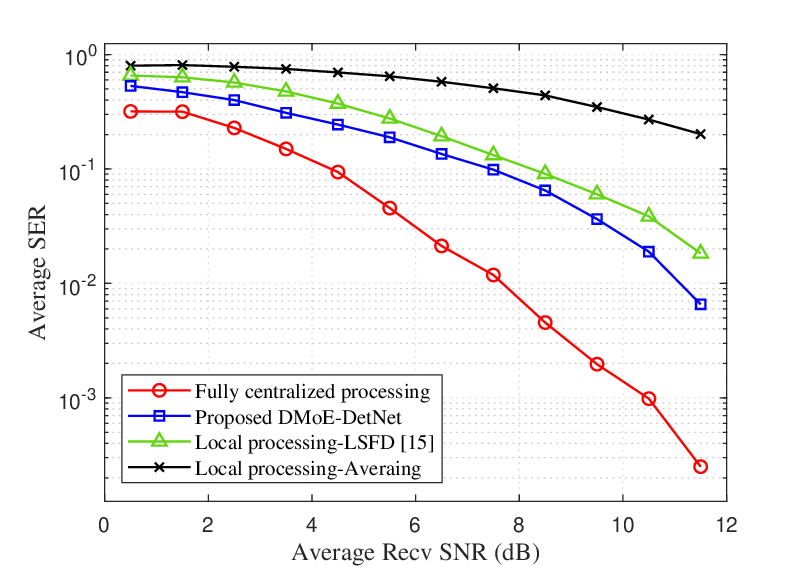}
		\caption{Average SER against the average receiver SNR ($K=8$, $Ls=4$, $N=8, 16$--QAM).}
		\label{fig:SER_vs_SNR}
	\end{figure}

	\begin{figure}[!t]
		\centering
		\includegraphics[width=0.955\columnwidth]{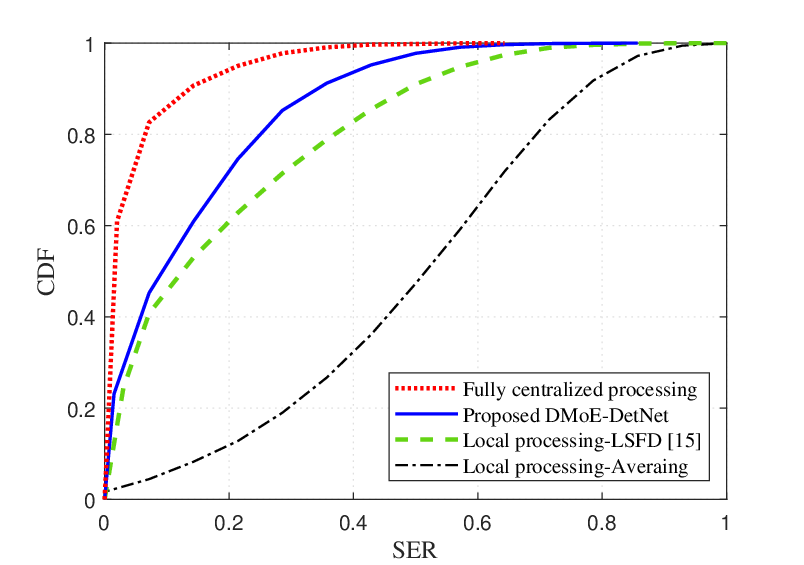} 
		\caption{CDF of average SER ($K=8$, $L=4$, $N=8, 16$-QAM).}
		\label{fig:SER_CDF}
	\end{figure}

	\myreffig{fig:SER_vs_SNR} illustrates the SER versus average received SNR across all AP-UE links. As SNR increases, all schemes exhibit performance improvements. Fully centralized processing achieves the optimal performance, while local processing with averaging combining yields the poorest results. Although LSFD weighting, optimized through statistical expectation, significantly mitigates inter-cell interference compared to equal weighting, its performance remains limited by reliance on channel statistical assumptions. Notably, proposed DMoE-DetNet, while sharing the same architectural framework as local processing with LSFD combining, demonstrates superior performance by adaptively learning from deterministic channel conditions and leveraging enhanced feature fusion and encoding, closely approaching the performance upper bound set by fully centralized processing.

	\myreffig{fig:SER_CDF} shows the cumulative distribution function (CDF) of SER, revealing trends consistent with \myreffig{fig:SER_vs_SNR}. DMoE-DetNet consistently outperforms local processing with LSFD combining, despite utilizing identical CSI and received signal information, which clearly validates the effectiveness and advancement of the proposed method.

	\begin{figure}[!t]
		\centering
		\includegraphics[width=0.955\columnwidth]{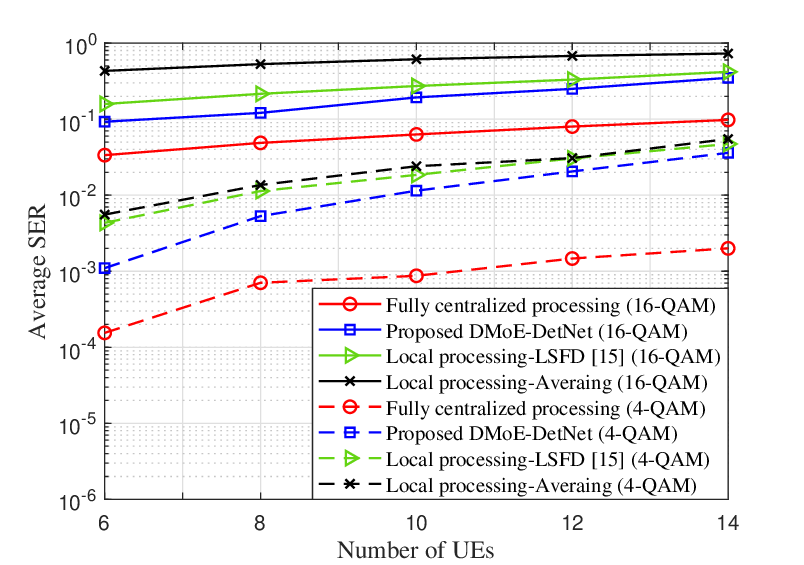} 
		\caption{Average SER against the number of UEs ($L = 4, N = 8, \{4 / 16\}$--QAM).}
		\label{fig:SER_UE}
	\end{figure}

	In \myreffig{fig:SER_UE}, we investigate the average SER versus the number of UEs under $4$--QAM and $16$--QAM modulation. As the number of UEs increases, SER monotonically rises due to intensified multi-user interference. The SER for $4$--QAM remains lower than that of $16$--QAM, benefiting from its lower modulation order. While fully centralized processing combining and local processing with averaging combining exhibit relatively stable performance degradation, both DMoE-DetNet and local processing with LSFD combining show accelerated performance degradation with more UEs, indicating higher sensitivity to UE contamination.

	\begin{figure}[!t]
		\centering
		\includegraphics[width=0.955\columnwidth]{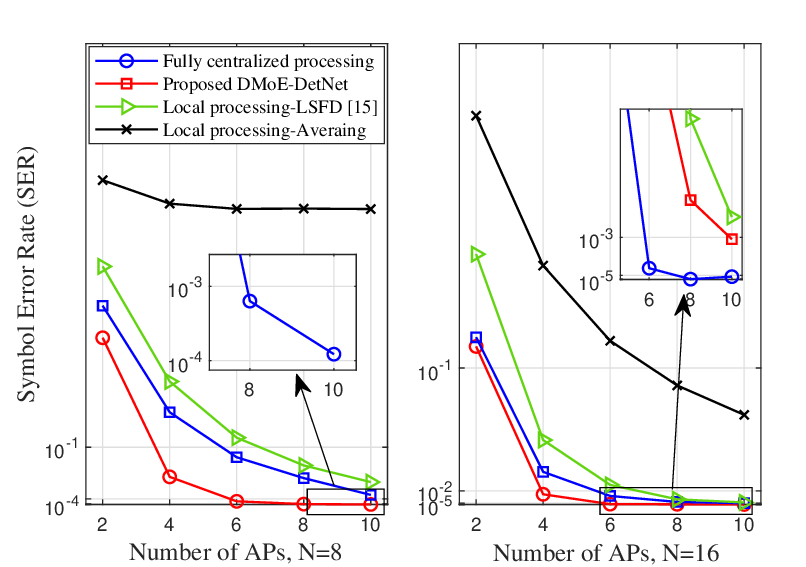} 
		\caption{Average SER against the number of APs and antennas ($K = 8, 16$--QAM).}
		\label{fig:SER_AP}
	\end{figure}

	\myreffig{fig:SER_AP} compares the SER performance of different methods under varying numbers of APs and antennas per AP. The SER monotonically decreases as the number of APs increases, due to enhanced macro diversity. Similarly, a larger number of antennas per AP also reduces SER. It is also observed that the rate of SER improvement gradually saturates as the number of APs grows.

	\section{Conclusion}
	\label{sec:conclusion}

	In this paper, we propose DMoE-DetNet, a novel deep learning framework for distributed uplink detection in cell-free massive MIMO systems. The architecture employs local expert networks at APs to extract non-linear features from ICSI, while a transformer-based encoder at the CPU captures global spatio-temporal dependencies. The key is the gating network that dynamically weights AP contributions using SCSI, enabling intelligent fusion.
	Comprehensive simulations demonstrate that our approach outperforms LSFD-based linear detectors, achieving near-optimal performance while maintaining practical fronthaul requirements. Compared to the fully centralized processing combing scheme that attains the maximum best SER, the proposed method can achieve comparable performance, with much reduced interaction overhead.

	\section{Acknowledgement}
	This work was supported in part by the National Natural Science Foundation of China under Grant NSFC 62301032, and Grant NSFC 62371040.

	\bibliographystyle{IEEEtran}
	\bibliography{IEEEabrv,Ref}

\end{document}